\documentclass[]{jfm}

\usepackage{graphicx}
\usepackage{newtxtext}
\usepackage{newtxmath}
\usepackage{natbib}
\usepackage[hidelinks]{hyperref}
\hypersetup{
    colorlinks = true,
    urlcolor   = blue,
    citecolor  = blue
}
\usepackage[justification=justified,
   format=plain,width=\textwidth]{caption}

\newcommand{\RomanNumeralCaps}[1]
\linenumbers

\graphicspath{{FIGURES/}}
\usepackage{amsmath}
\usepackage[inkscapeformat=png]{svg}
\usepackage{calrsfs}
\DeclareMathOperator{\pe}{\mathit{Pe}}
\DeclareMathOperator{\pr}{\mathit{Pr}}

\DeclareMathOperator{\nus}{\mathit{Nu}}

\DeclareMathOperator{\ra}{\mathit{Ra}}

\usepackage{color}

\title{Transport scaling in porous media convection}

\author{
Xiaojue Zhu\aff{1}\corresp{\email{zhux@mps.mpg.de}},
Yifeng Fu \aff{1}
\and
Marco De Paoli\aff{2,3}\corresp{\email{m.depaoli@utwente.nl}}
}

\affiliation{\aff{1}Max Planck Institute for Solar System Research, 37077 G\"ottingen, Germany
\aff{2}Physics of Fluids Group and Max Planck Center for Complex Fluid Dynamics and J. M. Burgers Centre for Fluid Dynamics, University of Twente, P.O. Box 217, 7500AE Enschede, The Netherlands
\aff{3}Institute of Fluid Mechanics and Heat Transfer, TU Wien, 1060 Vienna, Austria
}

\begin{document}
\maketitle

\begin{abstract}
We present a theory to describe the Nusselt number ($\nus$), corresponding to the heat or mass flux, as a function of the Rayleigh--Darcy number ($\ra$), the ratio of buoyant driving force over diffusive dissipation, in convective porous media flows. 
First, we derive exact relationships within the system for the kinetic energy and the thermal dissipation rate. 
Second, by segregating the thermal dissipation rate into contributions from the boundary layer and the bulk, which is inspired by the ideas of the Grossmann and Lohse theory (\emph{J. Fluid Mech.}, vol. 407, 2000; \emph{Phys. Rev. Lett.}, vol. 86, 2001), we derive the scaling relation for $\nus$ as a function of $\ra$ and provide a robust theoretical explanation to the empirical relations proposed in previous studies. Specifically, by incorporating the length scale of the flow structure into the theory, we demonstrate why heat or mass transport differs between two-dimensional and three-dimensional porous media convection. 
Our model is in excellent agreement with the data obtained from numerical simulations, affirming its validity and predictive capabilities.
\end{abstract}


\section{Introduction}
Carbon dioxide (CO$_2$) sequestration is a process aimed at long-term storage of large volumes of CO$_2$ \citep{schrag2007preparing}, primarily to mitigate climate change and support energy transition.
One of the most promising sequestration strategies involves natural underground reservoirs. 
In this case, liquid CO$_2$ is injected in saline aquifers, geological porous formations located hundreds of meters beneath the Earth's surface and confined by horizontal impermeable layers \citep{hup14,depaoli2021influence}. 
Saline aquifers are filled with brine, highly salted water denser than CO$_2$.
Because of this density difference, the injected volume of CO$_2$ will sit on top of the brine, and a critical configuration takes place: in case of fractures in the top confining layer of the formation, CO$_2$ would migrate upwards and eventually reach the upper strata up to the atmosphere \citep{hidalgo2015dissolution}.
However, CO$_2$ is partially soluble in brine and the resulting mixture, which is heavier than both starting fluids, sinks downward through the porous rocks and makes CO$_2$ safely trapped \citep{emami2015convective,letelier2023scaling}. 
To determine the optimal CO$_2$ injection rate and predict the long-time behavior of the injected CO$_2$, it is therefore imperative to conduct a meticulous assessment of the flow dynamics and the associated mixing laws \citep{macminn2012spreading,guo2021novel}. 
An idealized representation of this complex system consists of a porous domain fully saturated with fluid and confined between a heated bottom plate and a cooled top plate \citep{hew12,wen2018rayleigh}.
The top-to-bottom temperature difference induces a density gradient that drives the flow.
This configuration, which we label here as Rayleigh--Darcy (RD) convection, serves as a fundamental model of the aforementioned process, where the CO$_2$ concentration field, responsible for the density increase in the case of geological carbon sequestration, is replaced by a temperature field. 
Indeed, thermal and solutal convective porous media flows  can be considered equivalent and controlled by the same governing equations provided that: (i) in the temperature-driven flow, the solid phase is locally in thermal equilibrium with the liquid phase, and (ii) in the corresponding concentration-driven system, the solid is impermeable to the solute.
Additional factors to be accounted for a proper comparison between these systems are the dependency of the viscosity and the fluid density with respect to the value of the scalar. 
While viscosity is generally weakly sensitive to temperature variations, it may be considerably affected by the local value of solute concentration. 
However, it has been previously shown by \citet{hidalgo2012scaling} that, in convective porous media flows, concentration-induced viscosity variations do not significantly affect the global transport properties of the system.
In contrast, the shape of the density-concentration (or density-temperature) curves is shown to be crucial. 
For a general introduction to the RD convection, we refer to the reviews by \cite{hew20} and \cite{pao23}.
In this work, we will refer to fluid characterized by a constant viscosity and a linear dependency of density with the transported scalar (temperature).

The single control parameter in RD convection is the Rayleigh--Darcy number $\ra$, which indicates the relative strength between driving forces (convection) and dissipative effects (diffusion and viscosity), while the major response parameter of the system is the Nusselt number $\nus$, a dimensionless measure of the amount of heat (or solute) exchanged. 
Similar to the Rayleigh-B\'enard (RB) convection (i.e., a fluid heated from below and cooled from the top, in the absence of any porous medium), in recent years, major efforts have been put into understanding the scaling relation between $\nus$ and $\ra$, where $\ra$ is intended as the thermal Rayleigh number \citep{ahl09}.
The classical theory \citep{pri54,mal54,how66} posits that at significantly high $\ra$, the buoyancy flux should be independent of the layer's height ($L$). 
In the high-$\ra$ regime within a porous medium, this argument predicts a linear scaling of $\nus\sim \ra$. 
It has also been rigorously demonstrated that the linear scaling serves as an upper bound \citep{doering1998bounds,ote04,wen12,has14}. 
Interestingly, such scaling also means that the dimensional flux is independent of thermal diffusivity and, as a result, a realization of the scaling indicates the system reaches the so-called ultimate regime \citep{hew12,pir21}. 
In comparison, in RB convection, a similar argument leads to $\nus\sim \ra^{1/3}$ \citep{pri54,mal54,how66}, different from the diffusion-free ultimate scaling $\nus\sim \ra^{1/2}$ proposed by \cite{kra62} and \cite{spi63}. 
A detailed introduction to the scalings in RB is provided by \cite{ahl09}, \cite{chi12} and \cite{xia23}.

Direct numerical simulations (DNS) have been conducted in both two and three dimensions for RD convection to investigate heat transfer scaling at finite $\ra$. 
The two-dimensional DNS at high $\ra$ ($10^3 \le \ra \le 10^4$) by \cite{ote04} suggested a slightly sub-linear $\nus(\ra)$ scaling. 
Subsequent DNS, as reported by \cite{hew12}, extended up to $\ra=4\times 10^4$, and indicated that the scaling $\nus\sim \ra$ is asymptotically attained, albeit with a correction to the linear scaling. 
A simple fit, $\nus=0.0069\ra+2.75$, was proposed to accommodate the data within this range. 
For three-dimensional RD convection, DNS conducted by \cite{pir21} and \cite{pao22} reached up to $\ra=8\times10^4$ and suggested that the appropriate scaling for $\nus$ is given by $\nus=0.0081\ra+0.067\ra^{0.61}$. 
This contrasts with an alternative fit proposed by \cite{hew14}, where $\nus=0.0096\ra+4.6$ was considered for the data within the range up to $\ra=2\times10^4$.

It is crucial to underline that the previously mentioned corrections to the linear scaling are purely empirical in nature. 
This leads to the fundamental question: can we provide an explanation for these corrections and quantify them? 
To address this, we turn our attention to the Grossmann-Lohse (GL) theory \citep{gro00,gro01}, a key tool for comprehending the effective scaling of Nusselt and Reynolds numbers in relation to $\ra$ in turbulent RB convection. 
The central premise of the GL theory can be summarized as follows: Firstly, it establishes a connection between $\nus$ and $\ra$ by considering their relationship with the kinetic energy dissipation rate and thermal dissipation rate through exact relations. 
Secondly, the theory dissects these dissipation rates into contributions from the boundary layer and the bulk flow. 
In this work, we will derive corresponding exact relations for RD convection. 
By applying the principles of the GL theory, we can deduce the boundary layer and bulk contributions to the thermal dissipation rate, shedding light on the origins and expressions of the corrections to the linear scaling.

\section{Governing equations} \label{sec:eqs}
\begin{figure}
\includegraphics[width=0.99\columnwidth]{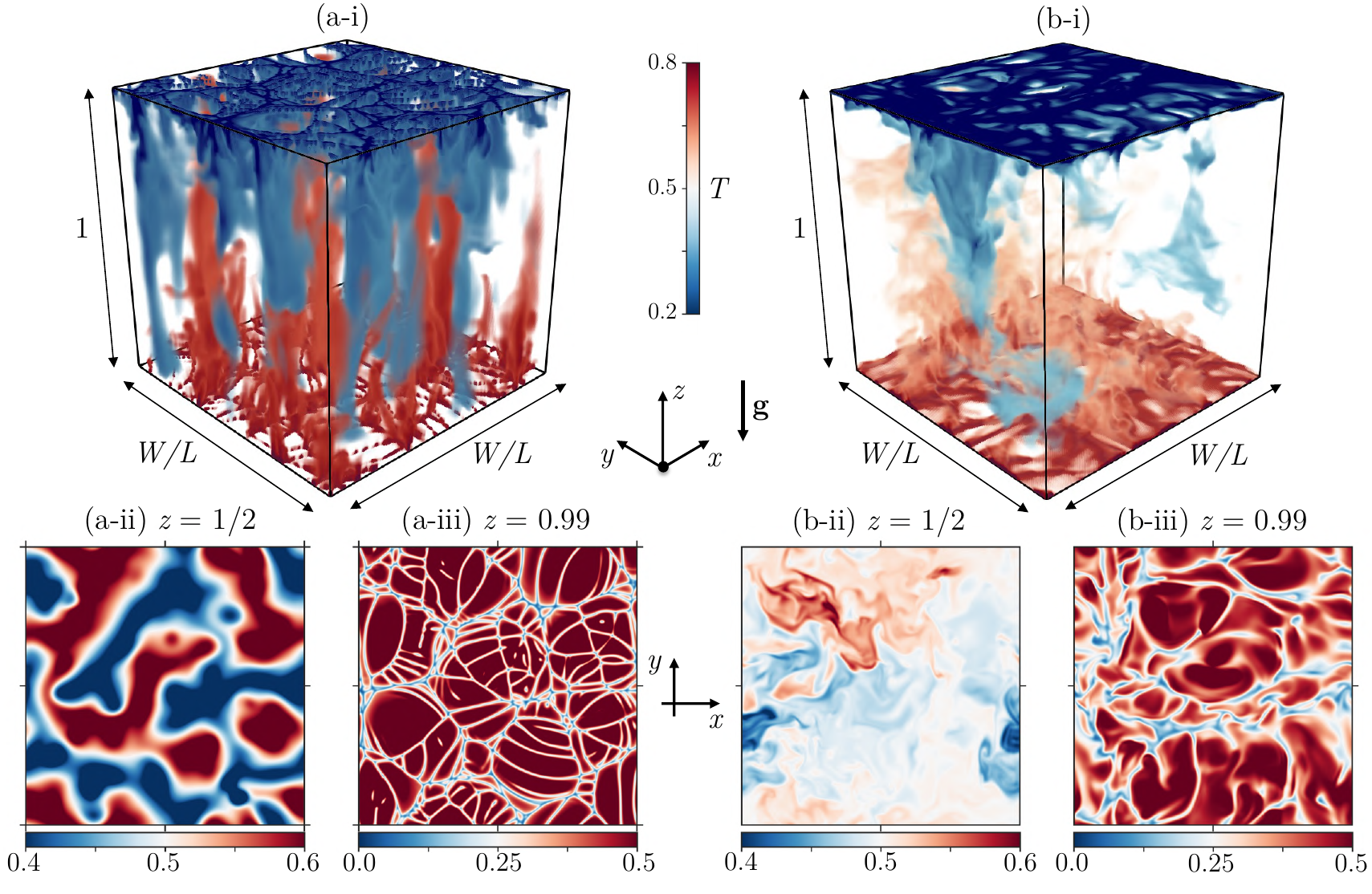}
\caption{Instantaneous dimensionless temperature field $T$ for convection in (a) porous media for $\ra=10^4$, $W/L=1$ \citep{pir21}, labeled as Rayleigh-Darcy (RD) convection, and (b)~in classical Rayleigh-B\'enard turbulence (thermal Rayleigh number $10^9$, Prandtl number $1$, $W/L=1$), labeled as RB. The temperature distribution is shown over the entire volume (a-i, b-i), at the centerline $z=1/2$ (a-ii, b-ii) and near the upper wall (a-iii, b-iii).}
\label{fig:fields} 
\end{figure}

We consider a fluid-saturated porous domain heated from below and cooled from above, as sketched in Figure~\ref{fig:fields}(a-i). 
Although we discuss here the problem of a thermally-driven flow in a porous medium that is locally in thermal equilibrium with the fluid, the same conclusions apply when the scalar is a solute, provided that the governing parameter ($\ra$) is matched.
The size of the domain considered in $W$ is the wall-parallel directions $x,y$, and $L$ in wall-normal direction $z$, along which gravity ($\mathbf{g}$) acts.
The flow is visualized in terms of dimensionless temperature $T$.
For sufficiently high $\ra$, a columnar flow structure develops both in 2D and in 3D, as one can observe from the cross section relative to the domain mid height in Figure~\ref{fig:fields}(a-ii), while the near-wall region (Figure~\ref{fig:fields}a-iii) is populated by thin filamentary plumes.
This structure differs significantly from the classical RB turbulence, reported in Figure~\ref{fig:fields}(b), which is controlled in the bulk by large-scale rolls. 
In RB convection, large-scale structures span the entire domain, with typical length scales comparable to the height of the system. In comparison, in RD convection, columnar-like structures prevail, and are characterized by length scales distinct from the system’s height. 
This difference will be taken into account when we build up a theory for RD convection in the following sections.

Before presenting the dimensionless equations, we introduce the scaling quantities employed. 
The equations are made dimensionless with respect to convective flow scales \citep{pir21}, namely, velocities are scaled with $\mathcal{U}=\alpha g\Delta K/\nu$, where $\alpha$ is the thermal expansion rate, $\Delta$ the temperature difference between the bottom and the top plate, $g$ the acceleration due to gravity, $\nu$ the kinematic viscosity and $K$ the permeability of the porous medium, which we assume to be homogeneous and isotropic.
Lengths are scaled with $L$ and time with $\phi L/\mathcal{U}$. 
The dimensionless temperature is obtained as $T=(T^*-T^*_\text{top})/\Delta$, being $T^*$ and $T^*_\text{top}$ the dimensional temperature field and the temperature value at the top boundary, respectively.
Finally, pressure is scaled by $g L (\rho^*_\text{top}-\rho^*_\text{bot})$, where the top-to-bottom fluid density difference ($\rho^*_\text{top}-\rho^*_\text{bot}$) is used.
In an incompressible RD system, the heat transport is controlled by the dimensionless advection-diffusion equation \citep{pir21}:
\begin{eqnarray}
    \frac{\partial T}{\partial t}+\mathbf{u}\cdot \nabla T&=&\frac{1}{\ra}\nabla^2T \text{  ,}
    \label{eq:01}    
\end{eqnarray}
where $\mathbf{u}$ and $T$ are the velocity and temperature fields, respectively, $t$ is time and $\ra$ is the Rayleigh-Darcy number defined as
\begin{equation}
    \ra=\frac{\alpha g\Delta KL}{\kappa\nu}.
    \label{eq:02}    
\end{equation}
In this parameter, the medium ($K$), domain ($g,L$) and fluid ($\alpha,\Delta,\nu,\kappa$) properties are included, where $\kappa$ is the thermal diffusivity.
The momentum transport and the flow incompressibility are accounted by the Darcy law and continuity equations, respectively:
\begin{equation}
    \mathbf{u}=-(\nabla p-T\mathbf{k})
    \label{eq:03}    
\end{equation}
\begin{equation}
    \nabla\cdot \mathbf{u}=0,
    \label{eq:04}
\end{equation}
where $p$ is the reduced pressure field and $\mathbf{k}$ is the unit vector aligned with $z$.

At the horizontal boundaries, we consider the temperature constant and equal to $T=1$ at the bottom plate and $T=0$ at the top, so that an unstable configuration is achieved and the flow is driven by convection. 
No-penetration boundary conditions are assumed at both plates for the velocity, while the sides are considered periodic. 
Equations~\eqref{eq:01}, \eqref{eq:03} and \eqref{eq:04} together with these boundary conditions determine the flow dynamics, which is controlled by two dimensionless parameters, namely $\ra$ and the horizontal domain width $W/L$.
The latter does not appear explicitly in the equations, but may play a significant role in determining the flow structure, especially at low $\ra$.

\section{Nusselt number and exact conservation equations}\label{sec:rel}
First, we will correlate the thermal (Nusselt number) and the kinetic (Péclet number) response parameters to the control parameter (Rayleigh-Darcy number), and then the thermal dissipation will be linked to the Nusselt number. 

\begin{figure}
\centering
\includegraphics[height=0.33\columnwidth]{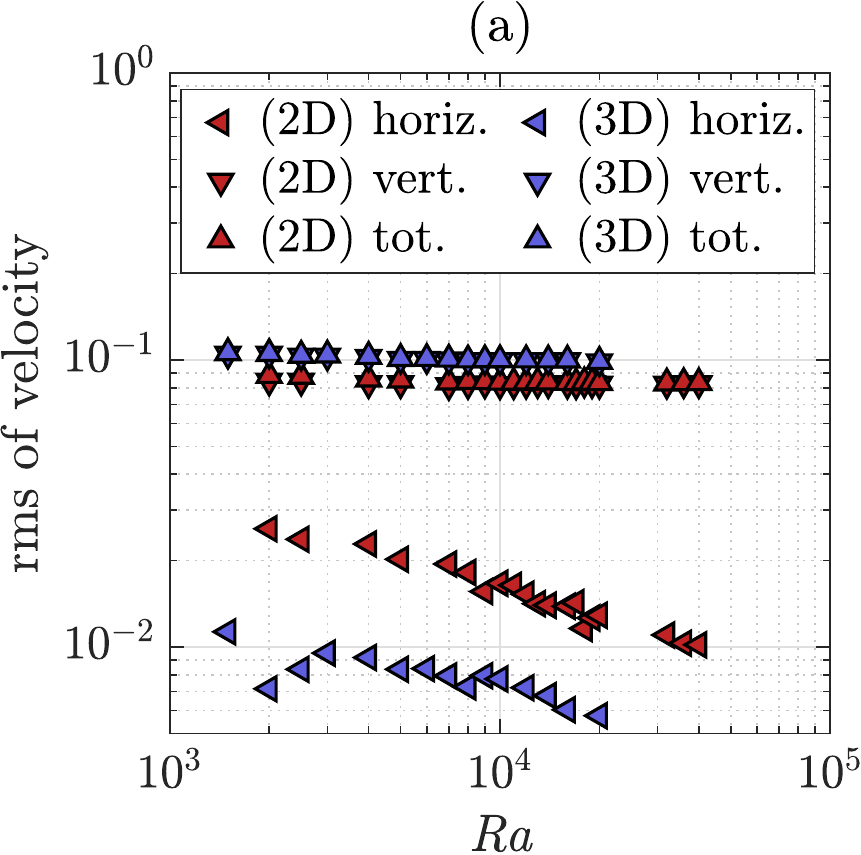}
\includegraphics[height=0.33\columnwidth]{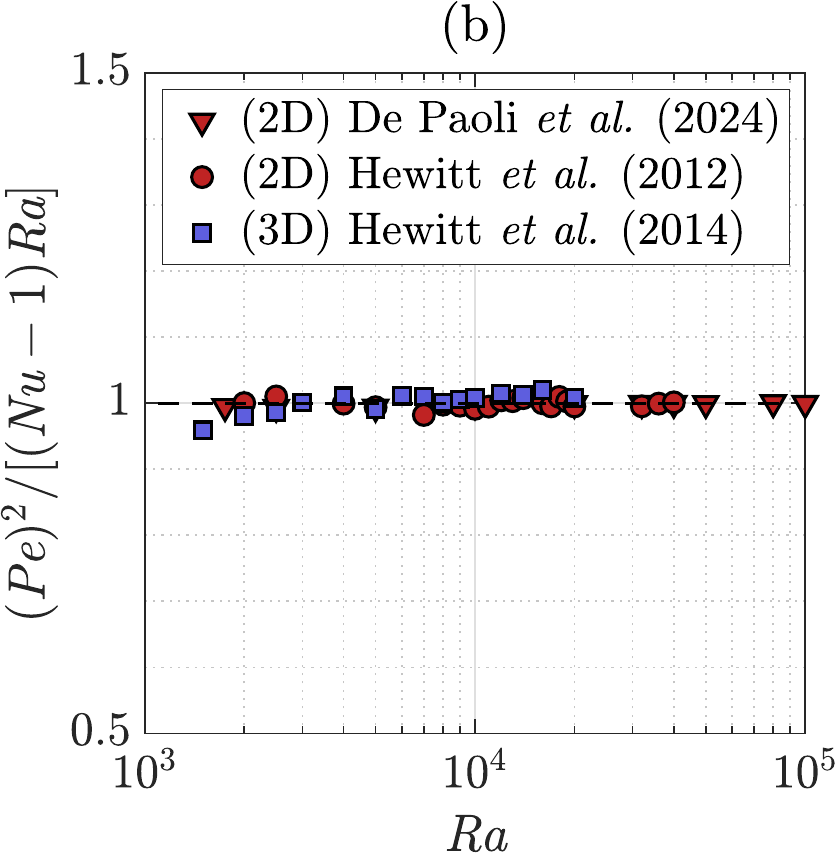}
\includegraphics[height=0.33\columnwidth]{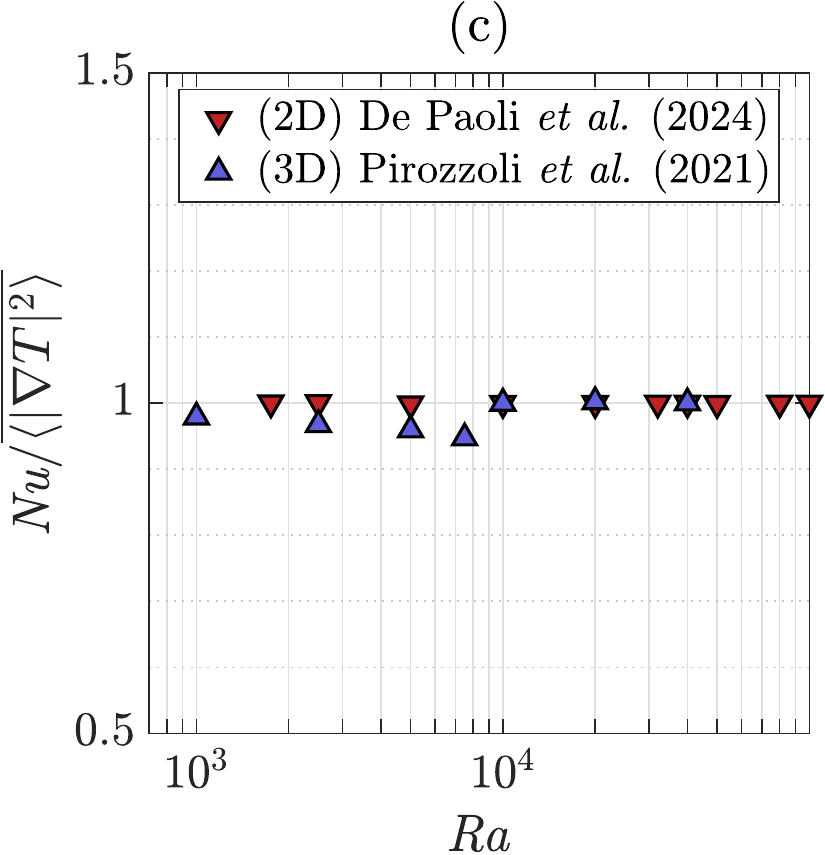}
\caption{
(a)~Root Mean Square (rms) of velocity used to estimate the Péclet number, $\pe$. 
Horizontal component ($\textnormal{rms}(u)$), vertical component ($\textnormal{rms}(w)$) and total (defined as $\sqrt{[\textnormal{rms}(u)]^2+[\textnormal{rms}(w)]^2}$ for 2D, and $\sqrt{2[\textnormal{rms}(u)]^2+[\textnormal{rms}(w)]^2}$ for 3D) are reported. 
Data from \cite{hew12,hew14}.
(b)~Comparison of $\pe$ compensated with $(\nus-1)\ra$ as from~\eqref{eq:pe2}.
$\pe$ is computed with $\mathcal{V}$ defined as in~\eqref{vpeb} for the data from \cite{hew12,hew14}, and as in Eq.~\eqref{eq:def_vel} for data from \citet{depaoli2024}.
(c)~Ratio of the Nusselt number to the dimensionless thermal dissipation, derived in~\eqref{eq:nuth}.
Data from \cite{pir21} and \citet{depaoli2024}. 
}\label{fig:rms} 
\end{figure}

The temporal and horizontal average of \eqref{eq:01} can be written as
\begin{equation}
    \frac{\partial}{\partial z} \left ( \ra \overline{\left \langle u_{z} T \right \rangle}_{A} - \overline{\left \langle \frac{\partial T}{\partial z} \right \rangle}_{A} \right ) = 0.
\end{equation}
The Nusselt number then reads \citep{letelier2019perturbative,ulloa2022energetics}
\begin{equation}
    \nus=\ra \overline{\left<u_zT\right>}_A-\overline{\left< \frac{\partial T}{\partial z} \right>}_A.
    \label{eq:Nu definition}
\end{equation} 
Here the following notations are used for different averaging procedures. 
Overbars $\overline{\cdots}$ correspond to the time average of a dimensionless value $f$, while an average over the horizontal surface and an average over the whole volume domain are denoted by $\left< \cdots\right>_A$ and $\left< \cdots\right>$, respectively:
\begin{eqnarray}
    \overline{f} &=& \frac{1}{\tau} \int_{t_0}^{t_0+\tau} f \mathrm{d} t\\ 
    \left \langle f \right \rangle_{A} &=& \frac{1}{A} \int_{0}^{W/L} \int_{0}^{W/L} f \mathrm{d}x \mathrm{d}y\\
    \left \langle f \right \rangle &=& \frac{1}{V} \int_{0}^{W/L} \int_{0}^{W/L} \int_{0}^{1} f \mathrm{d}x \mathrm{d}y \mathrm{d}z,
\end{eqnarray}
where $A=(W/L)^2$ is the dimensionless horizontal surface area and $V=(W/L)^2$ is the dimensionless volume of the whole domain based on our characteristic length scale $L$. 
Two exact relations exist in our system and can be derived from the governing equations.
Using the dimensionless velocity $\mathbf{u}$ to dot product both sides of \eqref{eq:03} and combining with the incompressible continuity equation \eqref{eq:04}, we get:
\begin{equation}
    \left | \mathbf{u} \right |^2 = -\nabla\cdot (p\mathbf{u}) + Tu_z.
    \label{eq:dimensionaless kinetic energy}
\end{equation} 
The volume and time average of \eqref{eq:dimensionaless kinetic energy} reads
\begin{equation}
    \overline{\left<  \left | \mathbf{u} \right |^2 \right>} = -\frac{1}{V} \iint_{\Sigma} (p\mathbf{u}) \cdot \hat{\mathbf{n}} \mathrm{d}S + \overline{\left< T u_z \right>},
    \label{eq:mean power equation}
\end{equation}
where $\Sigma$ is the boundary surface of the domain, $\mathrm{d}S$ denotes the surface element on the boundary, and $\hat{\mathbf{n}}$ is the normal unit vector for the surface elements. The mean power given by pressure gradient vanishes due to the non-penetration boundary condition:
\begin{equation}
    \frac{1}{V} \iint_{\Sigma} (p\mathbf{u}) \cdot \hat{\mathbf{n}} \mathrm{d}S = -\frac{1}{V} \iint_{\Sigma(z=0)} p u_z \mathrm{d}S  +  \frac{1}{V} \iint_{\Sigma(z=1)} p u_z \mathrm{d}S = 0.
    \label{eq:pressure gradient power}
\end{equation}
The mean buoyancy power in \eqref{eq:mean power equation} can be written as
\begin{equation}
    \overline{\left< T u_z \right>} = \frac{1}{V} \int_{0}^{1} \overline{\left < Tu_z \right >}_A A \mathrm{d}z = \frac{1}{\ra} \int_{0}^{1} \left (\nus + \overline{\left< \frac{\partial T}{\partial z} \right>}_A \right ) \mathrm{d}z.
    \label{eq:buoyancy power 1}
\end{equation} 
Here the last equivalence comes from the $\nus$ definition \eqref{eq:Nu definition}. 
Since we use $L$ as our length scale, $z\in[0,1]$. The last term in the above equation reads:
\begin{equation}
    \frac{1}{\ra}\int_{0}^{1} \overline{\left< \frac{\partial T}{\partial z} \right>}_A  \mathrm{d}z = \frac{1}{\ra}\int_{0}^{1} \frac{\partial \overline{\left< T \right>}_A}{\partial z} \mathrm{d}z = \frac{1}{\ra}\left ( \overline{\left< T \right>}_A |_{z=1} - \overline{\left< T \right>}_A |_{z=0} \right) = -\frac{1}{\ra}.
    \label{eq:integral}
\end{equation} 
Reintroducing \eqref{eq:integral} back into \eqref{eq:buoyancy power 1}, after some algebraic manipulations we get:
\begin{equation}
    \overline{\left< T u_z \right>} =  \frac{1}{\ra} \int_{0}^{1} \left (\nus + \overline{\left< \frac{\partial T}{\partial z} \right>}_A \right ) \mathrm{d}z = \frac{1}{\ra} (\nus - 1).
    \label{eq:buoyancy power 2}
\end{equation}
Combining \eqref{eq:mean power equation}, \eqref{eq:pressure gradient power} and \eqref{eq:buoyancy power 2}, we obtain an expression for the mean dimensionless velocity square:
\begin{equation}
    \overline{\left<  \left | \mathbf{u} \right |^2 \right>} = \frac{1}{\ra} (\nus - 1).
    \label{eq:mean u^2}
\end{equation}
We introduce the velocity scale 
\begin{equation}
\mathcal{V} = \mathcal{U}\sqrt{\overline{\left <\left | \mathbf{u} \right |^2 \right>}}, 
    \label{eq:def_vel}
\end{equation}
with $\mathcal{U}=\alpha g\Delta K/\nu$, and one finally obtains an exact relation:
\begin{equation}
    \pe^2=(\nus-1)\ra
    \label{eq:pe2}
\end{equation}
with
\begin{equation}
\pe = \frac{\mathcal{V} L}{ \kappa} = \ra\frac{\mathcal{V}}{\mathcal{U}},
    \label{eq:pe3}
\end{equation}
where $\pe$ is the Péclet number. The relation \eqref{eq:pe2} aligns with the findings reported by \cite{has14}, albeit derived from a slightly different set of equations for porous media convection. Note that for RB convection, the analogous exact relation is $\epsilon_u=\nu^3/L^4(\nus-1)\ra\pr^{-2}$, where $\epsilon_u$ is the kinetic energy dissipation rate and $Pr$ is the Prandtl number \citep{ahl09}.
To assess the validity of Eq.~\eqref{eq:pe2}, we consider the numerical measurements available in literature.
For 2D flows, $\pe$ is measured by \citet{depaoli2024} using \eqref{eq:def_vel} and \eqref{eq:pe3}.
The velocity Root Mean Square (rms) at the centerline (2D and 3D) is computed by \citet{hew12,hew14} and reported in Figure~\ref{fig:rms}(a).
Since in all directions no mean flow exists, we have that $\mathcal{V}$ is obtained from the rms of the velocity components ($u_i$), namely:
\begin{equation}
    \mathcal{V} = \mathcal{U}\left(\overline{\left <[\text{rms}(u_x)]^2+[\text{rms}(u_y)]^2+[\text{rms}(u_z)]^2\right >}\right)^{1/2}.\label{vpea}
\end{equation}
Assuming that the centerline flow is representative of the kinetic energy of the system, we have that:
\begin{eqnarray}
    \mathcal{V} \approx\mathcal{U}\left(\overline{\left <[\text{rms}(u_x)]^2+[\text{rms}(u_y)]^2+[\text{rms}(u_z)]^2\right >_{\Sigma(z=1/2)}}\right)^{1/2}.
    \label{vpeb}
\end{eqnarray}
We use this approximation to compute $\mathcal{V}$ and verify the validity of \eqref{eq:pe2} for the data of \citet{hew12,hew14}.
 We finally observe in Figure~\ref{fig:rms}(b) that Eq.~\eqref{eq:pe2} (dashed line) is in excellent agreement with the measurements obtained from the exact definition of $\pe$ \citep[2D and $\mathcal{V}$ computed with Eq.~\eqref{eq:def_vel} from][]{depaoli2024} and also with measurements obtained from the approximated definition of $\pe$ \citep[2D and 3D, $\mathcal{V}$ computed with Eq.~\eqref{vpeb} from][]{hew12,hew14}.

We will now derive an equation to correlate the mean thermal dissipation to the Nusselt number. 
Multiplying the dimensionless thermal advection-diffusion equation \eqref{eq:01} by $T$, integrating over the whole domain and time-averaging, we have:
\begin{equation}
    \overline{\left < \frac{\partial}{\partial t} \left ( \frac{T^2}{2} \right ) \right >} = - \overline{\left < T \mathbf{u} \cdot \nabla T \right >} + \frac{1}{\ra} \overline{\left < T \nabla^2 T \right >}.
    \label{eq:T2 equation}
\end{equation} 
Under the assumption of statistically steady state:
\begin{equation}
     \overline{\left < \frac{\partial}{\partial t} \left ( \frac{T^2}{2} \right ) \right >} = \frac{\partial}{\partial t} \overline{\left < \frac{T^2}{2} \right >} = 0.
     \label{eq:T2 1}
\end{equation}
The first term on the right hand side of \eqref{eq:T2 equation} can be further written as
\begin{equation}
    -\overline{\left < T \mathbf{u} \cdot \nabla T \right >} = \overline{\left < \frac{T^2}{2} \nabla \cdot \mathbf{u} \right >} - \overline{\left < \nabla \cdot \left ( \mathbf{u}\frac{T^2}{2}  \right) \right >} = -\frac{1}{V} \iint_{\Sigma} \left( \mathbf{u}\frac{T^2}{2} \right) \cdot \hat{\mathbf{n}} \mathrm{d}S = 0.
    \label{eq:T2 2}
\end{equation}
Here we employed again continuity \eqref{eq:04} and the no-penetration boundary condition. 
The second term on the right hand side of \eqref{eq:T2 equation} reads
\begin{equation}
    \frac{1}{\ra} \overline{\left < T \nabla^2 T \right >} = \frac{1}{\ra} \left[ \overline{\left < \nabla^2 \left( \frac{T^2}{2} \right) \right >} - \overline{\left < \left\lvert \nabla T \right\lvert^2 \right >} \right].
    \label{eq:T2 3}
\end{equation}
Combining results from \eqref{eq:T2 equation}-\eqref{eq:T2 3}, one obtains
\begin{equation}
    \overline{\left < \left\lvert \nabla T \right\lvert^2 \right >} = \overline{\left < \nabla^2 \left( \frac{T^2}{2} \right) \right >}.
    \label{eq:T2 4}
\end{equation}
We can use the following procedure to further simplify the right hand side of \eqref{eq:T2 4}
\begin{eqnarray}
    \overline{\left < \nabla^2 \left( \frac{T^2}{2} \right) \right >} &=&  \frac{1}{V} \iint_{\Sigma} \left( \overline{T \nabla T} \right) \cdot \hat{\mathbf{n}} \mathrm{d}S\\ 
    &=& \frac{A}{V} \left ( \overline{\left< T\frac{\partial T}{\partial z} \right>}_{\Sigma(z=1)} - \overline{\left< T\frac{\partial T}{\partial z} \right>}_{\Sigma(z=0)} \right ) = \nus.
    \label{eq:T2 5}
\end{eqnarray}
Here we considered that $V=A=(W/L)^2$, applied the boundary conditions for $T$ and $u_z$, as well as the $\nus$ definition \eqref{eq:Nu definition}. 
Combining \eqref{eq:T2 4} and \eqref{eq:T2 5}, we get
\begin{equation}
    \overline{\left < \left\lvert \nabla T \right\lvert^2 \right >} = \nus,
    \label{eq:nuth}
\end{equation}
where $\overline{\left < \left\lvert \nabla T \right\lvert^2 \right >}$ represents the dimensionless mean thermal dissipation. 
This relation, which holds also for RB convection \citep{ahl09}, has been also presented before for RD flows \citep{ote04,hidalgo2012scaling,pao23} and for Hele-Shaw convection the limit of infinitely thin domains \citep{letelier2019perturbative,ulloa2022energetics}.
The ratio of the Nusselt number to the dimensionless thermal dissipation is compared in Figure~\ref{fig:rms}(c) for the numerical results of \cite{pir21} and \citet{depaoli2024}. 
We observe that, also in this case, the agreement between theory and simulations is good, confirming the validity of \eqref{eq:nuth}.
Finally, one obtains the dimensional thermal dissipation rate:
\begin{equation}
    \epsilon=\kappa\frac{\Delta^2}{L^2} \overline{\left < \left\lvert \nabla T \right\lvert^2 \right >} = \kappa\frac{\Delta^2}{L^2} \nus.
    \label{eq:ET}
\end{equation}
Equations \eqref{eq:pe2} and \eqref{eq:ET} represent the two exact relations we derived in our system.

\section{Application of GL theory and scaling relation for the Nusselt number}\label{sec:theo}

\begin{figure}
\centering
\hspace{0.4cm}
\includegraphics[height=0.33\columnwidth]{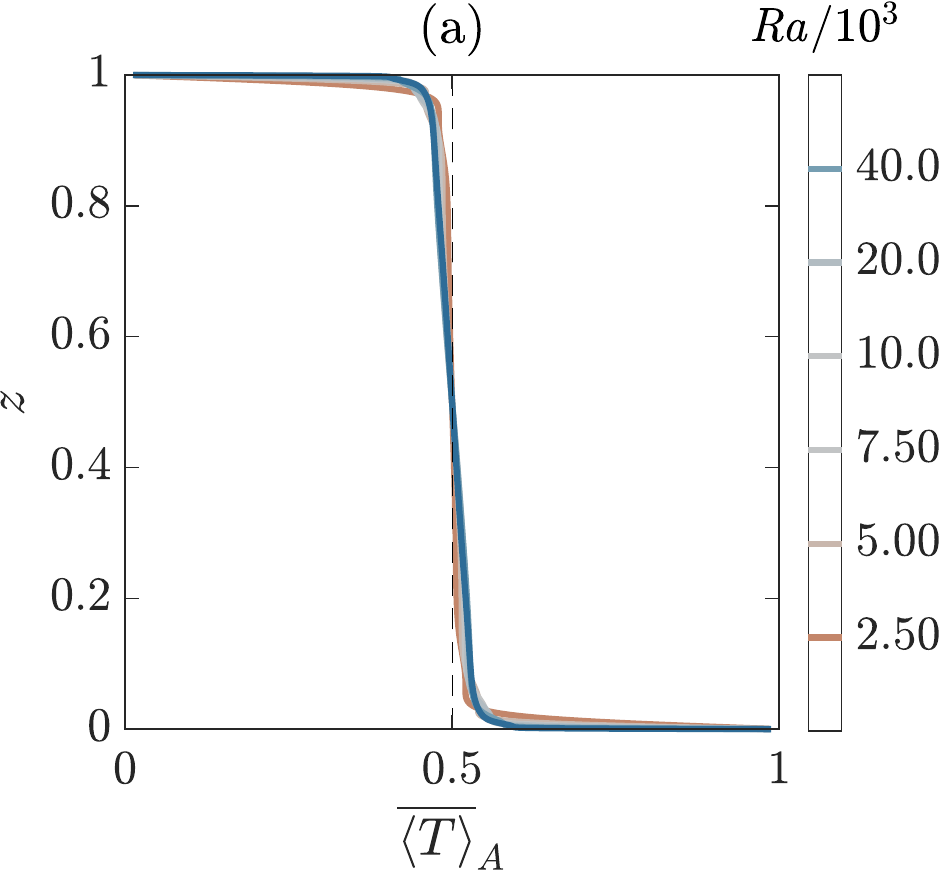}
\hspace{0.5cm}
\includegraphics[height=0.33\columnwidth]{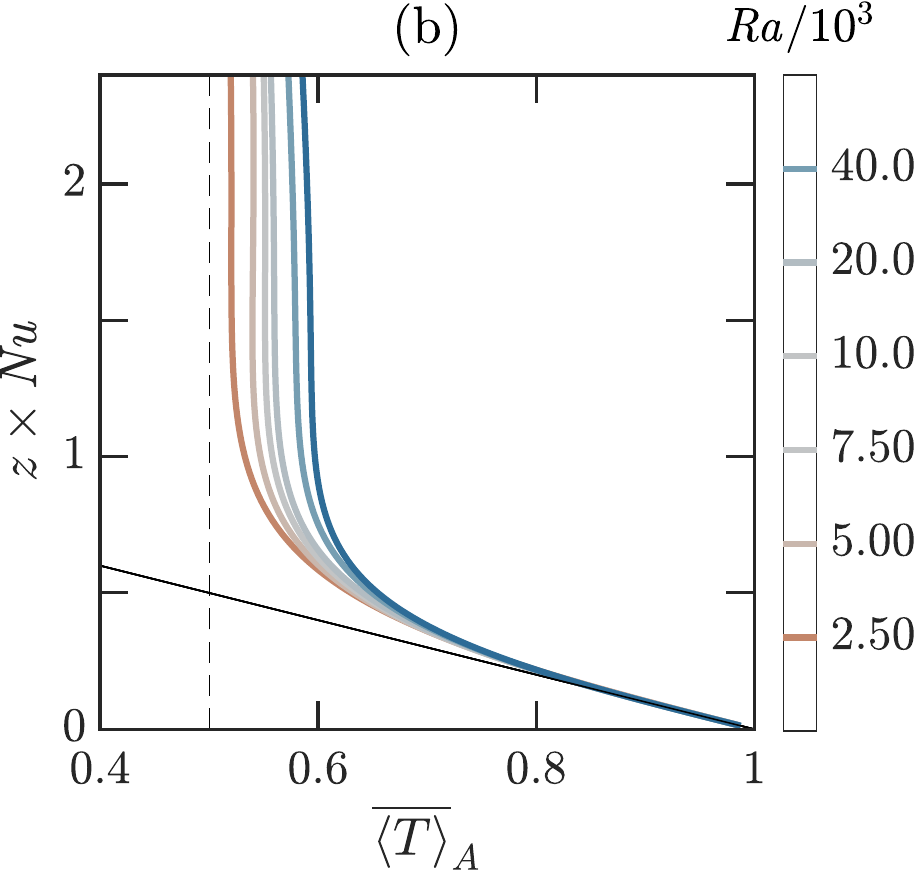}

\vspace{0.4cm}

\hspace{-0.15cm}
\includegraphics[height=0.33\columnwidth]{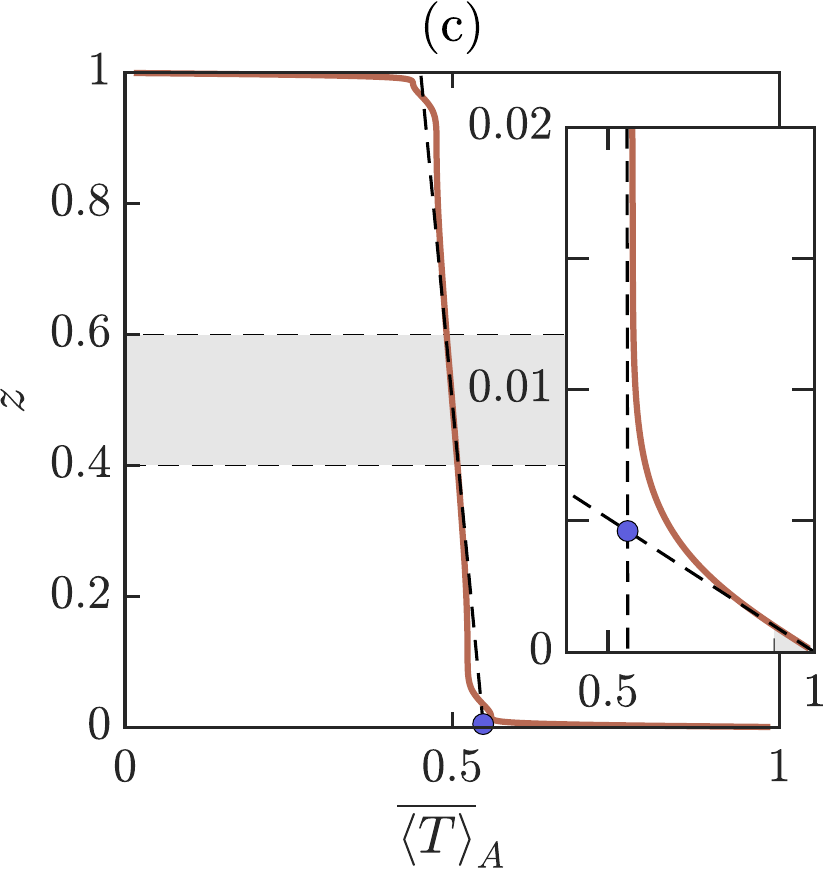}
\hspace{0.65cm}
\includegraphics[height=0.33\columnwidth]{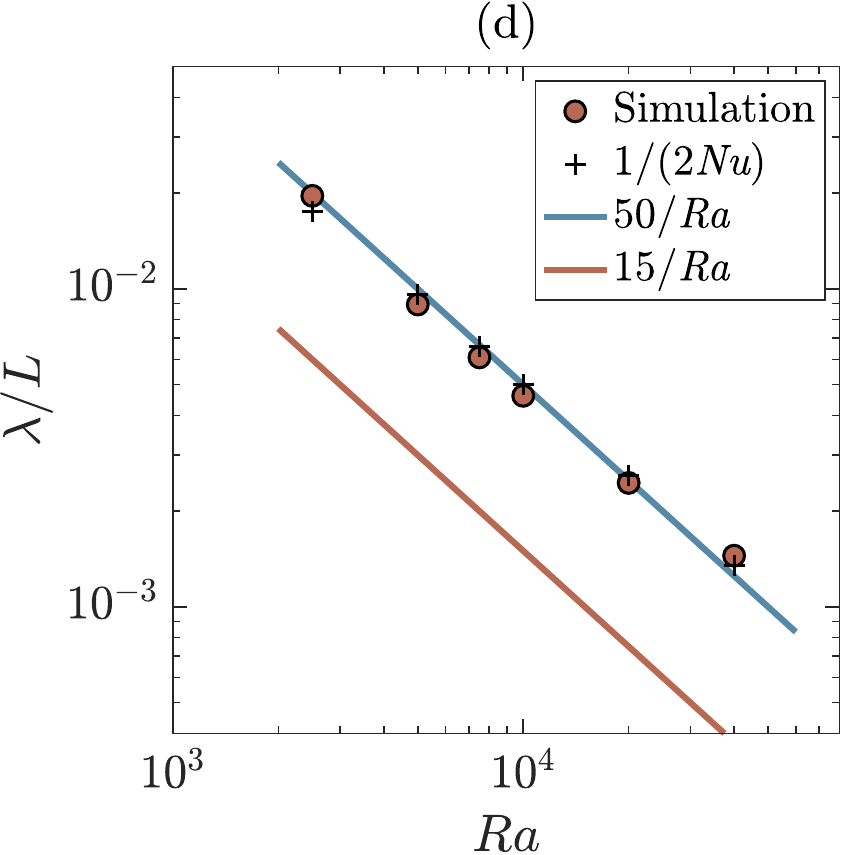}
\caption{
(a)~Horizontal- and time-averaged profiles of temperature are shown for different $\ra$ \citep{pao22}.
The near-wall region is magnified in (b), where the wall-normal coordinate is rescaled with $\nus$.
The profiles are self-similar, and in the boundary layer follow a linear behavior (solid line) with unitary slope.
(c)~The thickness of the boundary layer is determined as the distance from the wall of the intersection between the linear profile fitting the bulk ($0.4\le z \le 0.6$) and the near-wall regions ($z\le0.1\nus/\ra$). 
The case relative to $\ra=10^4$ is reported.
A close-up view of the near-wall region is shown in the inset.
(d)~Thickness of the thermal boundary layer $\lambda/L$ as a function of the Rayleigh number.
3D measurements computed as discussed (bullets) are very well fitted by the correlation $50/\ra$.
The correlation proposed for 2D flows by \cite{ote04} ($\lambda/L\sim15/\ra$) is also reported, as well as the value obtained from the Nusselt number $\lambda/L=1/(2\nus)$.
Data from \cite{pao22}.
}\label{fig:prof} 
\end{figure}

With the two exact relations derived in the previous section, we can now apply the main ideas of the GL theory to RD convection. 
The key idea of the GL theory \citep{gro00,gro01} is to split the kinetic and thermal dissipation rates into contributions from the boundary layers (BL) and bulk. 
In RD convection, the procedure becomes even simpler than in RB, as only the thermal dissipation rate appears in the exact relations. 
We separate the mean thermal dissipation as
\begin{equation}
    \epsilon=\epsilon_{BL}+\epsilon_{bulk},
    \label{eq:distot}    
\end{equation}
and apply the respective scaling relations for $\epsilon_{BL}$ and $\epsilon_{bulk}$, based on the boundary layer theory and fully developed flow in the bulk.
The horizontal- and time-averaged profiles of temperature, shown in Figure~\ref{fig:prof}(a), confirm that the flow can be split into two distinct regions: a well-mixed bulk with nearly-uniform properties, and a thin boundary layer characterized by a linear temperature profile, the slope of which is unitary when $z$ is rescaled with $\nus$ (Figure~\ref{fig:prof}b).
The thickness of this boundary layer, $\lambda/L$, can be determined as the distance from the wall at which the linear function fitting the temperature profile in the bulk ($0.4\le z \le 0.6$) intersects the near-wall temperature fit.
The measurement procedure is illustrated in Figure~\ref{fig:prof}(c), where the intersection is marked by the bullet.
The Nusselt number sets the thickness of the boundary layer $\lambda/L =1/(2\nus)$ \citep{pao22}.
In RD convection, it has been proposed by \cite{ote04} that the thermal boundary layer thickness scales as $\lambda/L \sim \ra^{-1}$ (consistent with $Nu \sim Ra$, from the classical theory \citep{pri54,mal54,how66} and the upper bound scaling derived by \citet{doering1998bounds}).
As illustrated in Figure~\ref{fig:prof}(d), this approximation is verified. 
Although both in 2D and 3D the thermal boundary layer thickness follows $\lambda/L \sim \ra^{-1}$, the prefactor differs (see Figure~\ref{fig:prof}d). 
This discrepancy arises from the distinct flow structures in 2D and 3D. 
In 3D, owing to the additional degree of freedom compared to the 2D case, plumes can freely move and reorganize towards the most efficient configuration, resulting in a different value of $\nus$. 
Consequently, the boundary layer thickness also varies. This phenomenon is analogous to RB convection, where 2D and 3D flows exhibit different boundary layer thicknesses due to variations in $\nus$ \citep{van2013comparison}.

\begin{figure}
\centering
\includegraphics[height=0.4\columnwidth]{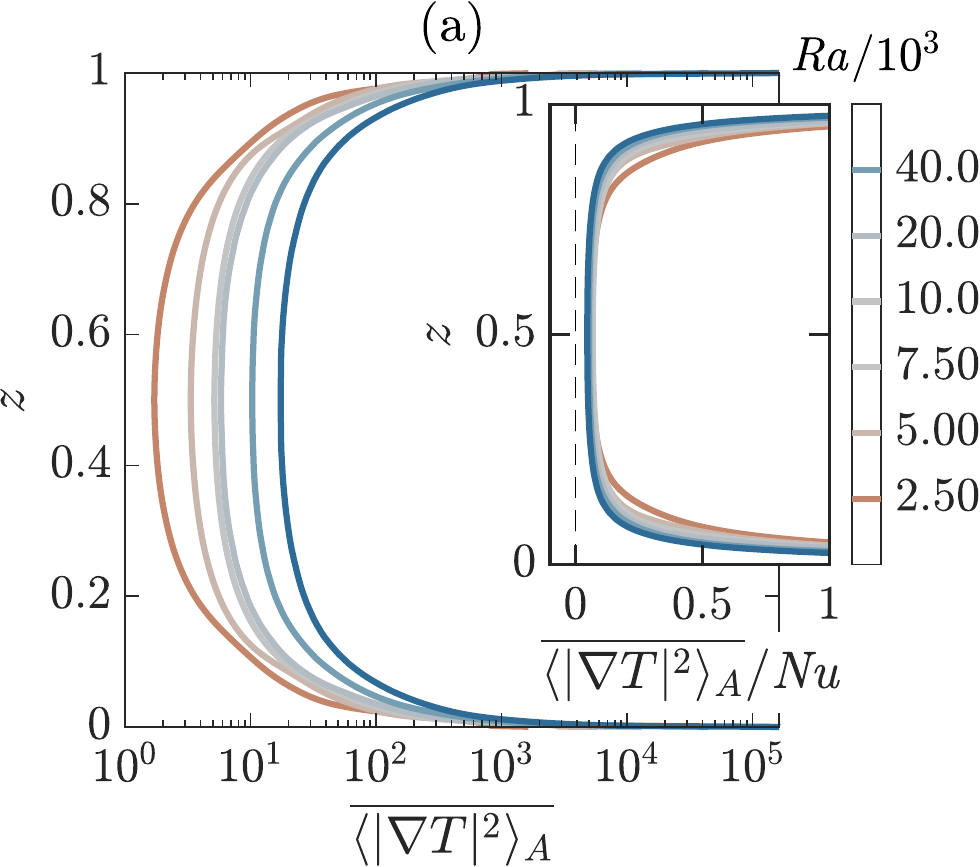}
\hspace{0.5cm}
\includegraphics[height=0.4\columnwidth]{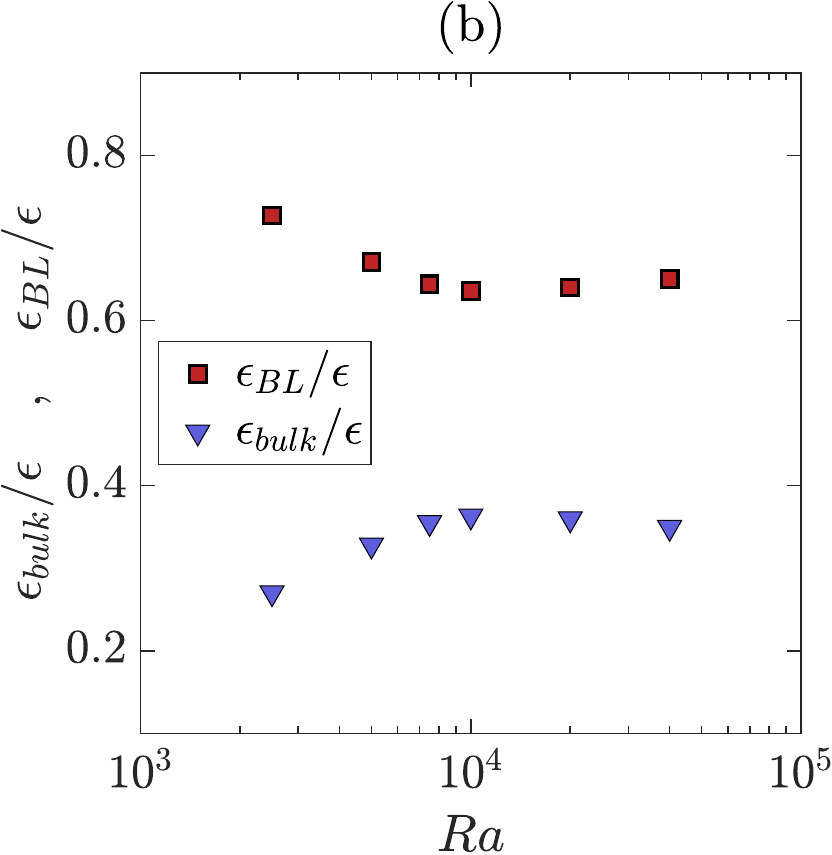}
\caption{
(a)~Horizontal- and time-averaged profiles of dimensionless thermal dissipation, $\overline{\langle|\nabla T|^2\rangle_A}$, are shown for different $\ra$ \citep{pao22}.
In the inset, the dissipation is rescaled by the respective Nusselt number, and shown up to $\overline{\langle|\nabla T|^2\rangle_A}/\nus=1$.
The boundary layer contribution to the dissipation is more pronounced as $\ra$ is increased. 
(b)~Contributions to the mean scalar dissipation $\epsilon$ within the boundary layer ($\epsilon_{BL}$) and in the bulk region ($\epsilon_{bulk}$). 
Data are from \cite{pao22}.
}\label{fig:bl} 
\end{figure}

The profiles of dimensionless thermal dissipation $\overline{\langle|\nabla T|^2\rangle_A}$ obtained from \cite{pao22} are shown in Figure~\ref{fig:bl}(a) for different values of the Rayleigh number. 
In the inset, the dissipation is rescaled by the respective Nusselt number, and shown up to 1.
We observe that the boundary layer contribution to the dissipation is more pronounced as $\ra$ is increased. 
A more quantitative description is provided in the following.
The thermal dissipation rate in the boundary layer scales as $\sim \kappa(\Delta/\lambda)^2$. 
Therefore, taking into account the layer extension ($\lambda/L\sim \ra^{-1}$), the boundary layer contribution towards the total thermal dissipation rate reads 
\begin{equation}
    \epsilon_{BL}\sim\kappa\frac{\Delta^2}{\lambda^2}\frac{\lambda}{L}\sim\kappa\frac{\Delta^2}{L^2}\ra.
    \label{eq:disBL}    
\end{equation}
Assuming the flow in the bulk is well mixed \citep*{gro00,bad23,son24}, we get 
\begin{equation}
    \epsilon_{bulk}\sim\frac{\mathcal{V}\theta^2}{\ell}.
    \label{eq:disbulk}
\end{equation}
Here, $\mathcal{V}$ and $\theta$ are the typical velocity and temperature scales, respectively.
The characteristic length scale is $\ell$, defined as the wavelength associated with the power-averaged mean wave number at the mid height $(\overline{k})$, i.e. $\ell/L=2\pi/\overline{k}$.
In numerical simulations, $\overline{k}$ is obtained from the time-averaged spectrum of the temperature field at $z=1/2$ \citep{hew14}.
The importance of $\epsilon_{BL}$ and $\epsilon_{bulk}$ relative to the total dissipation $\epsilon$, is reported in Figure~\ref{fig:bl}(b) for 3D RD convection \citep{pao22}.
Here, $\epsilon_{BL}$ and $\epsilon_{bulk}$ are obtained from the profiles, and represent the mean value within the respective regions.
In RB, to derive the scalings, the length scale is assumed to be the height of the domain $L$, which makes sense as there exist large-scale rolls. 
However, here in RD, the typical flow structures are columnar-like, making the length scale $\ell$ different from $L$. 
This difference is clearly visible in Figures~\ref{fig:fields}(a-i) and \ref{fig:fields}(b-i).
From the definition of $\nus$ (equation \ref{eq:Nu definition}) and assuming that in the bulk, $\theta \sim \nus(\kappa\Delta )/(\mathcal{V}L)$, and when $\nus$ only comes from the fluctuation in the bulk, $\theta \sim (\epsilon_{bulk}/\mathcal{V})( L/\Delta)$, we get
\begin{equation}
    \epsilon_{bulk}\sim\kappa\frac{\Delta^2}{L^2}{\pe}\frac{\ell}{L}.
    \label{eq:ebul10}
\end{equation}
The same bulk scaling has also been reported for rapidly rotating convection \citep*{son24} and magnetoconvection with strong vertical magnetic fields \citep{bad23}. In all these three systems, in the bulk there exists a new horizontal dominant length scale that is different from the height of the domain. In each of these three systems, a new dominant length scale emerges, distinct from the
domain’s height. This disparity constitutes a significant deviation from the original GL
theory.

\begin{figure}
\centering
\includegraphics[height=0.35\columnwidth]{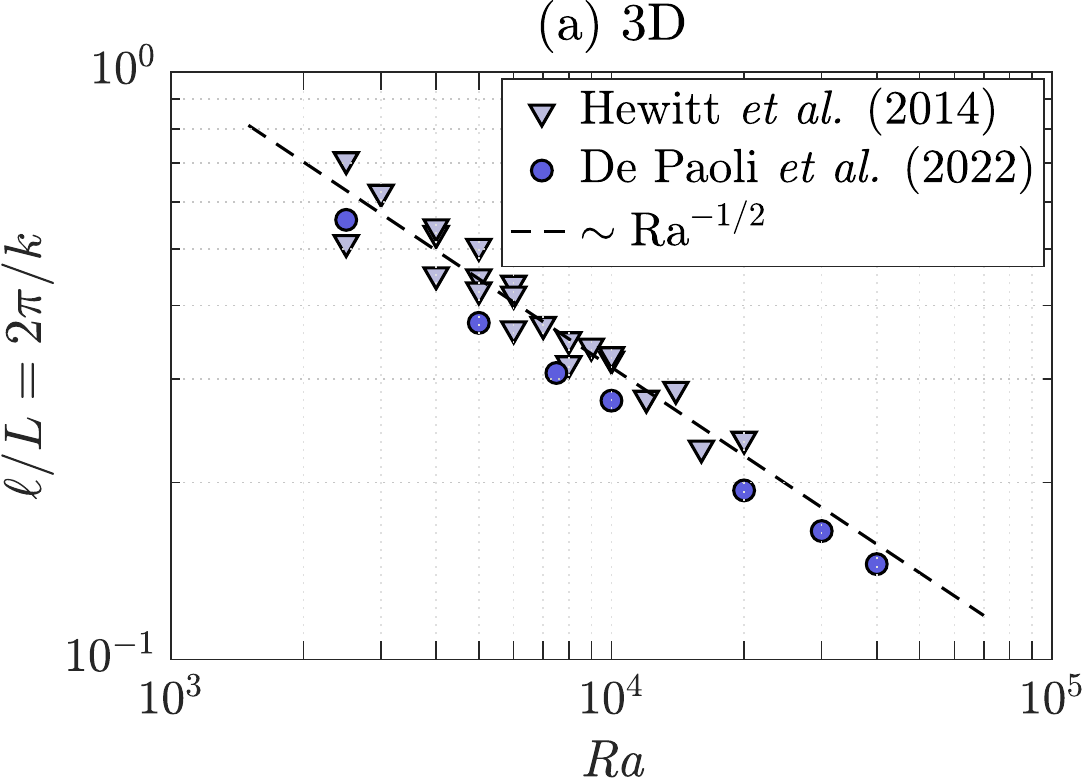}
\includegraphics[height=0.35\columnwidth]{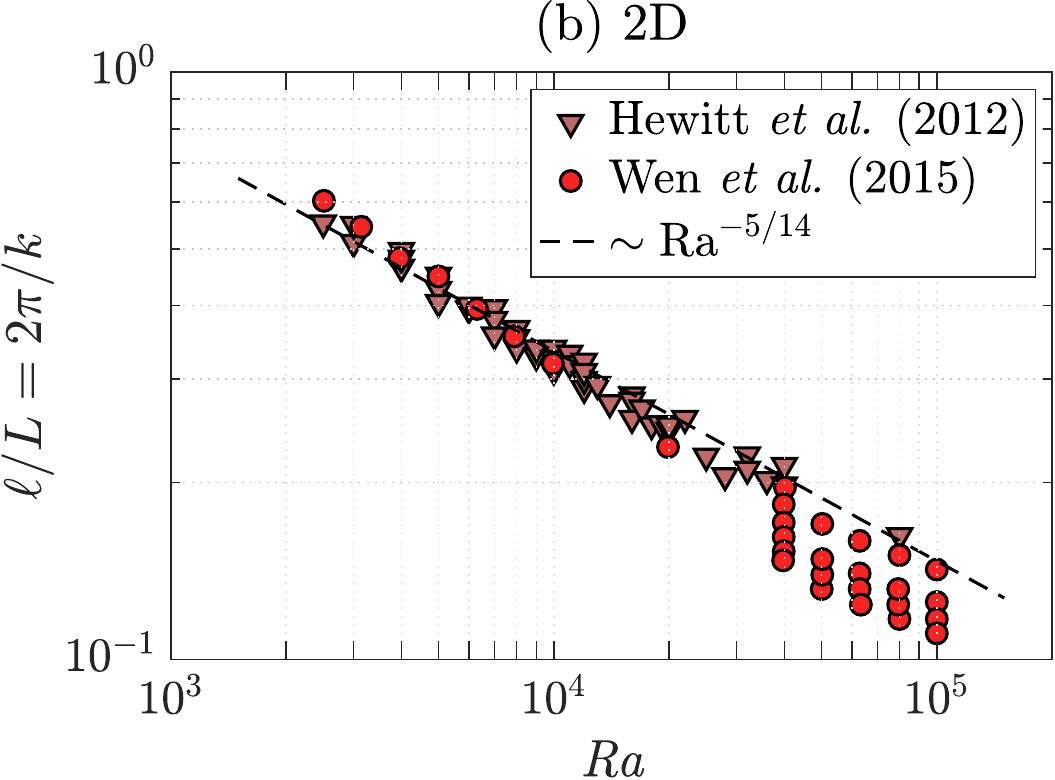}
\caption{
Dominant length scales $\ell/L$ in the bulk ($z=1/2$) for (a)~3D \citep{hew14,pao22} and (b)~2D \citep{hew12,wen2015structure} simulations. 
Theoretical scaling relations defined in Eq.~\eqref{eq:k3} and \eqref{eq:k2}, corresponding to 3D and 2D flows respectively, are also indicated (dashed lines).
}\label{fig:wn} 
\end{figure}

Determining the dominant wavelength in RD convection is a challenging task.
The reason is linked to the complex way in which the dynamic near-wall flow structures interact with the stationary, large-scale columnar plumes controlling the bulk. 
A detailed review is provided by \cite{hew20}, which we summarize here with additional details including later results \citep{pao22}. 
In 3D, using asymptotic stability theory, \cite{hewitt2017stability} derived that 
\begin{equation}
    \ell/L\sim \ra^{-1/2} \quad (3D).
    \label{eq:k3} 
\end{equation}
Numerical simulations by \cite{hew14} and \cite{pao22}, best fitted by scaling exponents of $-0.52$ and $-0.49$, respectively, agree well with this prediction (see also Figure~\ref{fig:wn}a). 
Therefore, we employ this scaling relation ($\ell/L\sim \ra^{-1/2}$) for the centerline in 3D. 
The situation is more complex in 2D. 
\citet{hewitt2013stability} derived analytically the scaling relation
\begin{equation}
    \ell/L\sim \ra^{-5/14} \quad (2D).
    \label{eq:k2}
\end{equation}
\cite{wen2015structure} have shown that for $\ra\le19976$ the centerline dominant length scale is well approximated by $\ell/L\sim \ra^{-0.40}$. 
However, one can observe in Figure~\ref{fig:wn}(b) that when $\ra\ge39716$, the inter-plume spacing measured by \cite{wen2015structure} is not unique.
The conclusion is that in 2D RD convection, at $\ra\ge39716$, a precise scaling remains to be established by running simulations in very wide domains and for very long times. 
In view of this, we consider the scaling proposed by \cite{hewitt2013stability}, which represents the best theoretical prediction available, to be valid.
Therefore, combining \eqref{eq:ebul10} with \eqref{eq:pe2}, \eqref{eq:k3} and \eqref{eq:k2}, we get 
\begin{equation}
\epsilon_{bulk}\sim\kappa\frac{\Delta^2}{L^2}\nus^{1/2}  \quad(3D).
    \label{eq:e3}
\end{equation}
and 
\begin{equation}
\epsilon_{bulk}\sim\kappa\frac{\Delta^2}{L^2}\nus^{1/2}\ra^{1/7}  \quad(2D).
    \label{eq:e2}
\end{equation}

\begin{figure}
\centering
\includegraphics[width=0.9\columnwidth]{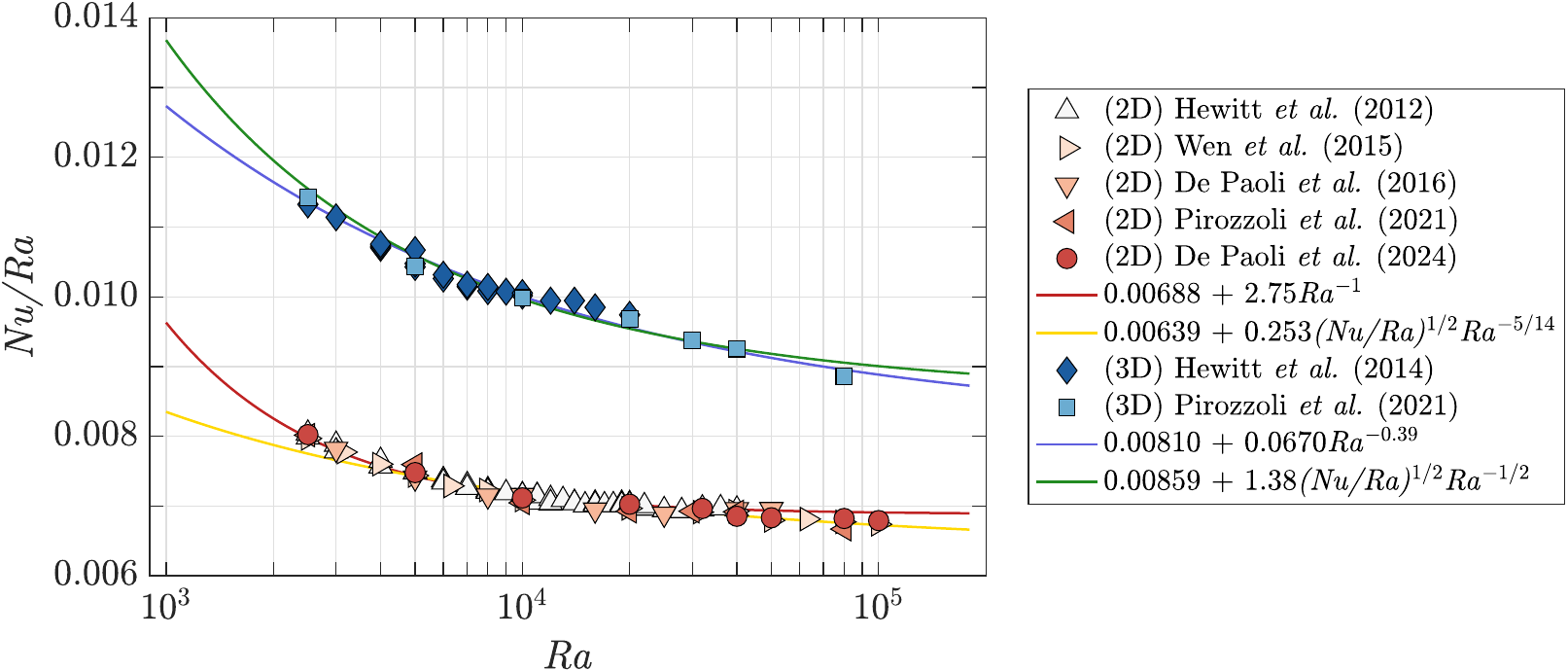}
\caption{
\label{fig:nu}
Compensated Nusselt number as a function of Rayleigh number.
Results are for 2D \citep{hew12,wen2015structure,depaoli2016influence,pir21,depaoli2024} and 3D \citep{pir21,hew14} simulations.
New fitting curves are obtained considering all data with $\ra>2\times10^3$, with numerical values of the coefficients being $A_2 = (6.386 \pm 0.007)\times10^{-3}$, $B_2=0.2533\pm0.0050$ and $A_3 = (8.592 \pm 0.033)\times10^{-3}$, $B_3=1.376\pm0.059$. 
The coefficients of determination ($R^2$) for the best fitting curves provided are 0.991 and 0.999 in the 2D and in the 3D cases, respectively.
}
\end{figure}

Finally, combining \eqref{eq:ET}, \eqref{eq:distot}, \eqref{eq:disBL}, \eqref{eq:e3} and \eqref{eq:e2}, we reach an expression for $\nus$ as a function of $\ra$ for the 3D and the 2D cases:
\begin{equation}
   \nus=A_3\ra+B_3\nus^{1/2}  \quad(3D),
   \label{eq:nu3d}
\end{equation}
\begin{equation}
   \nus=A_2\ra+B_2\nus^{1/2}\ra^{1/7} \quad (2D).
   \label{eq:nu2d}   
\end{equation}
As reported in Figure~\ref{fig:nu}, these scaling relations fit very well the data obtained from numerical simulations, both in the 2D and in the 3D cases. 
The values of the coefficients $A_2$, $A_3$, $B_2$, $B_3$, indicated in Figure~\ref{fig:nu}, are obtained as best fitting from the data shown, representing the numerical results available and with $\ra>2\times10^3$.
The choice of considering values larger than this threshold is motivated by the flow topology: at low $\ra$ the bulk flow structure is not columnar, as it is dominated by large-scale convective rolls \citep{graham1994plume}, and therefore our theory does not apply. 
The expressions of $\nus$ derived in \eqref{eq:nu3d} and \eqref{eq:nu2d} take the form of a linear scaling with a sublinear correction. 
The linear scaling was previously proposed for 2D \citep{hew12} and 3D \citep{hew14} flows.
The scaling proposed here provides similar results to the linear scaling with sublinear corrections proposed for the 3D case by \cite{pir21}.
However, in this case one fitting parameters less is used, i.e., all scaling exponents are known. 
The good agreement of the present scaling relations with the numerical measurements available in literature suggests that the theory proposed is indeed valid and promising for higher $\ra$.
The RD system is completely defined by two parameters, namely the Rayleigh number $\ra$ and the aspect ratio $W/L$. 
In 3D flows at high-$\ra$, it has been observed that all major flow statistics converge for an aspect ratio of $W/L=1$ \citep{pao22}.
This differs in 2D systems: also at high $\ra$ and due to the additional lateral confinement, the aspect ratio may have an influence on $\ell$ \citep{wen2015structure}.
Therefore, it may be required to take $W/L$ into account in the present theory to better describe the transport properties in 2D RD flows.
To this aim and also to assess the physics of the scaling prefactors, additional simulations in large domains and at larger $\ra$ are needed.

\section{Conclusions}\label{sec:conc}
In summary, we have established two exact relationships, one pertaining to the Péclet number and the other to the thermal dissipation rate, in the context of Rayleigh-Darcy convection - a fundamental system for heat and mass transport in porous media.
Inspired by previous models developed for Rayleigh-Bénard convection \citep{gro00,gro01}, we have formulated a scaling theory for heat transfer in 2D and 3D Rayleigh-Darcy flows, where the Nusselt number is expressed as a function of the Rayleigh number as described by \eqref{eq:nu3d} and \eqref{eq:nu2d}. 
This theory enables us to provide a theoretical explanation to the sublinear empirical corrections proposed in prior studies \citep{pir21}. 
Our investigations, supported by both 2D and 3D literature results, confirm the validity of the proposed theory. 
 Moreover, by taking the length scale of the flow structures into account, we
 also shed new light on the physical origins of the disparities in scaling relations between 2D and 3D Rayleigh-Darcy convection.

Our findings are relevant to convective flows in homogeneous and isotropic porous media where the top-to-bottom density difference is defined. 
However, these hypotheses represent idealized conditions not taking into account additional flow features that occur in realistic processes, such as hydrodynamic dispersion \citep{liang2018,Tsinober2022,tsinober2023numerical}, medium heterogeneity \citep{dentz2023mixing,simmons2001variable}, anisotropy \citep{depaoli2016influence,ennis2005onset} and alternative flow configurations \citep{hidalgo2012scaling,depaoli2017solute,letelier2023scaling}.
Nonetheless, present findings represent a crucial step required to develop a robust and physically-sound theory for convection in porous media flows, which can be further expanded to include the presence of the different variations mentioned above.
We consider for instance the sequestration of carbon dioxide in saline aquifers.
Such a system is usually modelled as a rectangular domain initially filled with brine and confined by two horizontal low-permeability layers, and therefore it is assumed to be impermeable at the bottom boundary (no-flux) \citep{hup14}.
Here, the solute enters from the top, in correspondence of which the concentration of CO$_2$ is constant.
This flow configuration, defined as ``one-sided" \citep{hewitt2013convective} or ``semi-infinite" \citep{ennis2005onset}, is subject to a transient behaviour: the average CO$_2$ concentration within the system will increase over time, and it will be progressively harder to keep dissolving solute.  
In quantitative terms this means that, after a short initial phase in which dissolution increases due to the formation and growth of the fingers, the flux of solute through the top boundary will later reduce as a result of the saturation of the domain.
The dynamics of such a system has been thoroughly characterized \citep{slim2014solutal}, and it is shown to be quantitatively related to the dynamics observed in RD convection \citep{hewitt2013convective,depaoli2017solute}.
In order to describe the evolution of the one-sided system with a simple box-model, accurate predictions of the transport scaling in RD convection are essential. 
In presence of high-permeability formations as the Utsira Sand reservoir at Sleipner \citep{bickle2007modelling}, the Rayleigh-Darcy number may be as high as $6\times10^5$ \citep{hewitt2013convective}, which is beyond current numerical capabilities.
As a result, the theoretical results provided in our work will play a crucial role as a tool to determine the long-term evolution of flows in semi-infinite domains.

\backsection[Funding]{We gratefully acknowledge the financial support from the Max Planck Society, the German Research Foundation through grants 521319293 and 540422505, the Alexander von Humboldt Foundation, and the Daimler Benz Foundation. 
This project has also received funding from the European Union's Horizon Europe research and innovation programme under the Marie Sklodowska-Curie grant agreement No.~101062123.} 

\backsection[Declaration of interests]{The authors report no conflict of interest.}
\backsection[Author ORCID]{
\\Xiaojue Zhu, \href{https://orcid.org/0000-0002-7878-0655}{https://orcid.org/0000-0002-7878-0655};
\\Yifeng Fu, \href{https://orcid.org/0009-0006-6332-045X}{https://orcid.org/0009-0006-6332-045X};
\\Marco De Paoli, \href{https://orcid.org/0000-0002-4709-4185}{https://orcid.org/0000-0002-4709-4185};
}

\bibliographystyle{jfm}
\bibliography{bibliography}

\end{document}